\documentclass{article}
\usepackage{sprocl}
\usepackage{epsfig}
\def\etal{{\it et al.}}

\def\del{\partial}

\def\bra{\langle}
\def\ket{\rangle}

\def\beq{\begin{equation}}
\def\eeq{\end{equation}}
\def\beqa{\begin{eqnarray}}
\def\eeqa{\end{eqnarray}}
\def\etal{{\it et.al.}}
\def\lra{\leftrightarrow}
\def\gsim{{\scriptstyle\,\stackrel{\sim}{>}\,}}

\def\tbst{\vrule height2.5ex depth0ex width0pt}%

\begin{document}
\begin{flushright}
  UH-511-922-98 \\ December 5, 1998
\end{flushright}
\vskip 0.5in

\title{HADRONIC $B$ MESON DECAYS\\
     - Getting Ready for CP Violation -}

\author{HITOSHI YAMAMOTO}

\address{Department of Physics and Astronomy, The University of Hawaii,\\
Honolulu, HI 96822 USA\\E-mail: hitoshi@uhheph.phys.hawaii.edu}

\maketitle

\vspace{4cm}
\abstracts{We review recent results on hadronic B meson decays including
rare as well as non-suppressed modes. The main emphasis is on those channels
relevant to measurement of $CP$ violation in the coming $B$ factories.
After briefly describing flavor-tagged charm counting, we cover
$B\to DK^{(*)}$, $B\to$ charmless 2-body decays, inclusive $\eta'$
production, and final-state-interaction phases in $B\to\Psi K^*$ and
$B\to D^*\rho$.}

\vspace{2cm}
\begin{flushleft}
.\dotfill .
\end{flushleft}
\begin{center}
{Presented at The Fourth International Workshop on Particle
Physics Phenomenology, Kaohsiung, Tawain, June 1998, to appear in proceedings.} 
\end{center}
\newpage

\section{Flavor-tagged charm counting}

According to experiments,
the semileptonic branching ratio of $B$ mesons, $B_{s.l.}$,
and the number of charms generated in a $B$ meson
decay, $n_c$, are (the numbers are averages over $B^0$ and $B^+$)
\beq
   B_{s.l.} = 10.41 \pm 0.29\, \hbox{\%\,\cite{PDG}} \,,\quad
   n_c = 1.10 \pm 0.05\hbox{\,\,\cite{CLEO-DX}}\,,
\eeq
while the corresponding theoretical predictions are \cite{NeuSach}
\beq
   B_{s.l} = 
            \left\{
            \begin{array}{l}
               12.0 \pm 1.0 \% \\
               10.9 \pm 1.0 \%
            \end{array}
            \right.,\;
   n_c = 
            \left\{
            \begin{array}{l}
               1.20 \pm 0.06 \% \\
               1.21 \pm 0.06 \%
            \end{array}
            \right.,\;\hbox{ for }
   \mu = 
            \left\{
            \begin{array}{l}
               m_b \\
               m_b/2
            \end{array}
            \right. ,     
\eeq
where the errors are largely due to uncertainties in quark masses.
We see that the agreement between experiment and theory is within
the errors for $B_{s.l.}$ and about two sigmas for $n_c$; namely,
it is inconclusive at present. 

\begin{figure}[htb]
\vspace{-.3cm}
\centerline{\epsfig{figure=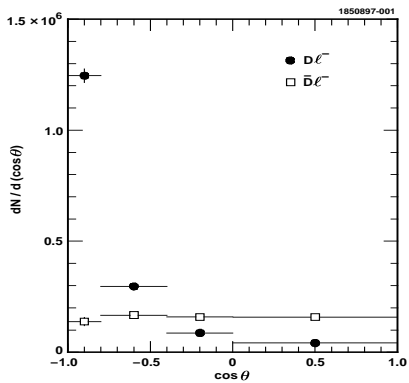,width=2.8in, height=2.2in}}
\vspace{-.3cm}
\caption{\label{fig:dlep}
   The $\cos\theta_{D{\rm-}\ell}$ distribution where $\theta_{D{\rm-}\ell}$
   is the angle between $D$ and the lepton. The charm quantum number of $D$
   and the charge of the lepton are the same for the
   open squares and opposite for the solid squares. If $D$ and the lepton
   are from the same $B$, they are mostly back-to-back.
   The lepton momentum is required to be greater than 1.5 GeV/c.}
\end{figure}

Further information can be obtained by
counting $c$ and $\bar c$ separately for a given $b$ flavor.
This is accomplished on the $\Upsilon 4S$ resonance~\cite{CLEOdlep} by
tagging $D^0$ and $D^+$ mesons (denoted
simply as $D$) with leptons from the other
$B$ meson. In the sample of $D$-lepton pairs, however,
$D$ and the lepton may come from the same $B$,
in which case $D$ and the lepton
tend to be back-to-back. Those coming from different $B$'s 
would be uncorrelated ($B$ mesons decay nearly at rest).
This can be used to statistically separate the two 
kinds of pairs.~\cite{lowPD}
Figure~\ref{fig:dlep} shows the measured angular correlations. 
One can see that the $\bar D\ell^-$ (or $D\ell^+$) pairs 
are mostly from different $B$'s, namely, 
the directions of $D$ and $\ell$
are uncorrelated (flat distribution). On the other hand,
the $D\ell^-$ (or $\bar D\ell^+$) pairs are mostly 
from the same $B$ (back-to-back); there is, however, a small uncorrelated
component. After corrected for the $B^0$-$\bar B^0$ mixing, the latter
component represents the `wrong-sign' $D$ meson 
production $b \to \bar D$ which
can be due to the process $b\to c W^-$, $W^-\to \bar c s$:
\beq
    {\#(\bar D\ell^+)_{opp. B} \over \#(D\ell^+)_{opp. B}} \to
   {Br(\bar B \to \bar D X)\over Br(\bar B \to D X)} =
   0.100\pm0.026\pm0.016\,.
\eeq
Before this measurement, $\bar c$ in $W^-\to \bar c s$ was usually
assumed to hadronize either as a $D_S^-$, one of its excited states, or a 
charmonium. A possibility of sizable wrong-sign $D$ production, however, 
had been pointed out prior to the measurement~\cite{BDY}.

As a byproduct, one can form the ratio of 
$\#(D\ell^+)_{opp. B}$, which measures $Br(\bar B\to DX)$, to 
$\#(D\ell^-)_{same\, B}$, which measures $Br(\bar B\to DX\ell^-\bar\nu)$,
to obtain
\beq
  r \equiv 
  { Br(\bar B\to DX)/Br(\bar B\to X) \over
    Br(\bar B\to DX\ell^-\bar\nu)/Br(\bar B\to X\ell^-\bar\nu) }
  = 0.901\pm0.034\pm0.015\,,
\eeq
where $Br(\bar B\to X)  = 1$ by definition, and 
$Br(\bar B \to X\ell^-\bar\nu)$ is the separately-measured
semileptonic branching ratio. Note that the numerator of $r$
is the probability that the $b$ quark will result in a right-sign
$D$, while the denominator is the same probability given that
the decay was semileptonic. If we ignore $b\to u$, $D_s^+$ production
from $b\to c$, production of charmed baryon or charmonium, and
$b\to \hbox{`$sg$'}$, then we expect $r=1$. 
Here, $b\to \hbox{`$sg$'}$ stands for any quark-level
transition $b\to$ (no charm) other than
$b\to u$. Correcting for all but
$b\to \hbox{`$sg$'}$, we get
\beq
   Br(b\to \hbox{`$sg$'}) = 0.2 \pm 3.4 \pm 1.5 \pm 1.7 \%\;
 \hbox{ or } <6.8\%\;(90\%\, c.l.)
\eeq
where the first error is statistical, second systematic, and the third
is due to models used in the corrections. It should be noted that
the ratio $r$ above is insensitive to the uncertainty in the
decay branching ratios of $D$ mesons used in the analysis since 
it largely cancels between the numerator and the denominator.

Another ratio of interest is the denominator of $r$ itself which is
expected to be unity if we ignore $\bar B\to D_s^+\ell^-\bar\nu$ and
$b\to u\ell^-\bar\nu$. The $D$ mesons are detected by $D^0\to K^-\pi^+$
and $D^+\to K^-\pi^+\pi^+$ where $Br(D^+\to K^-\pi^+\pi^+)$ is
normalized to $Br(D^0\to K^-\pi^+)$. Thus, by setting the above ratio to
unity after the corrections, one can extract $Br(D^0\to K^-\pi^+)$. 
The result is
\beq
  Br(D^0\to K^-\pi^+) = 3.69\pm0.11\pm0.16\pm0.04\,\%\,,
\eeq
which may be compared to the PDG value
$Br(D^0\to K^-\pi^+) = 3.85\pm0.09\,\%$~\cite{PDG}.

\section{$B^-\to D^0 K^{(*)-}$}

Along with $B\to K\pi,\pi\pi$,
these are modes that we hope can be used to measure the
CKM angle $\gamma$. A large CP asymmetry is possible in
$B^-\to D^0_i K^-$ vs. $B^+\to D^0_i K^+$ ($D^0_i$: CP eigensates)
through interference of $B^-\to D^0 K^-$ and $B^-\to \bar D^0 K^-$
(and its charge conjugate modes). The original method~\cite{GroWyl} 
proposed
to measure these 4 modes to construct two triangles to extract
the angle $\gamma$. A difficulty in measuring the `wrong-sign'
$D^0$ decay due to the doubly-Cabibbo-suppressed $D^0$ decay was
pointed out by Atwood, Dunietz and Soni,~\cite{AtDuSo} who also
proposed a solution for the problem even though it required
a large data sample. Analyses using only the favored modes were
recently proposed by Gronau~\cite{Gronau} and Xing.~\cite{Xing}
Also, the charged $B$ decays to $DK^{(*)}$ can be combined with
their corresponding neutral $B$ to enhance 
sensitivities.~\cite{JangKo,GroRos}

We detect $D^0$ in $B^-\to D^0 K^-$ in the modes
$D^0\to K^-\pi^+$, $K^-\pi^+\pi^0$, and $K^-\pi^+\pi^-\pi^+$.
The major background comes from the Cabibbo-favored counter part
$B^-\to D^0 \pi^-$. In fact, we expect
\beq
    { Br(B^-\to D^0 K^-) \over Br(B^-\to D^0 \pi^-) }\sim
    \lambda^2 \left({f_K\over f_\pi}\right)^2
    \sim 0.08\,,
\eeq
where $\lambda\sim 0.225$ is the Cabibbo factor and $f_{K,\pi}$ are
decay constants. The $B^-\to D^0 \pi^-$ background is suppressed
by the ionization loss information in the drift chamber and
the fact that miss-assigning kaon mass to the pion results in
the total energy greater than the expected $B$ meson energy.
There is also substantial background from the continuum
events, and they are reduced by cutting on the angle between
the sphericity axis of the $B$ candidate and that of the
rest of the event, the polar angle of $B$ in the lab frame, 
and a Fischer
discriminant. The Fischer discriminant $F$
is a linear combination of a set of measured variables $\vec x$
\beq
   F = \vec\lambda\cdot\vec x
\eeq
where the coefficients $\vec\lambda$ is determined by
maximizing the separation between the signal sample and the
background sample:
\beq
    S \equiv {(\bra F\ket_s - \bra F\ket_b)^2 \over \sigma_F^2}
    = {[ \vec\lambda\cdot(\bra\vec x\ket_s - \bra\vec x\ket_b)]^2
    \over \vec\lambda^T V \vec\lambda}\,.
\eeq
where $V$ is the covariant matrix of $\vec x$ and 
$\bra \;\ket_{s,b}$ denotes the average over the signal or
background sample.
Taking $\del S/\del \lambda_i = 0$ gives
\beq
   \vec \lambda = V^{-1} (\bra\vec x\ket_s - \bra\vec x\ket_b)\,.
\eeq
Here, $\vec x$ are the energy flows in 9 cones around the
event axis, the second Fox-Wolfram event shape variable, and
the polar angle in lab of $B$ candidate event axis.

\begin{figure}[htb]
\vspace{-.1cm}
\centerline{\epsfig{figure=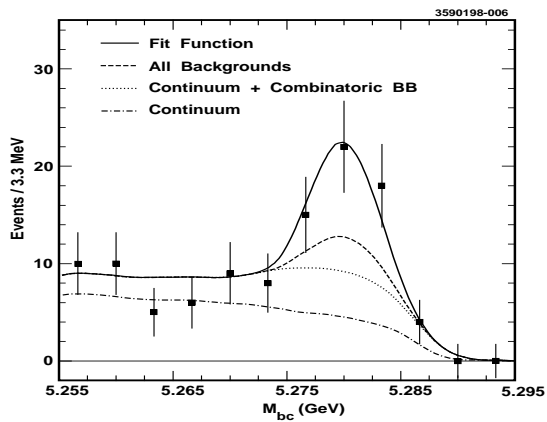,width=2.8in, height=2.2in}}
\vspace{-.3cm}
\caption{\label{fig:DK}
   The beam-constrained mass distribution for the $B^-\to D^0K^-$
   candidates. The events between the dotted line and the dashed line
   represent the background from $B^-\to D^0\pi^-$.}
\end{figure}

The $B$ mass is calculated from the total 3-momentum $\vec P_{tot}$ 
of the daughter particles and the known beam energy $E_{beam}$ as
\beq
    M_B \equiv \sqrt{E^2_{beam} - \vec P_{tot}^2}\,,
\eeq
and called the beam-constrained mass. Note that the
beam-constrained mass is equivalent to the absolute value of the
total momentum of candidate particles.
The distribution of $M_B$ after the background
rejection cuts is given in Figure~\ref{fig:DK}. 
There is a significant excess beyond the expected background and
we obtain~\cite{CLEODK}
\beq
    { Br(B^-\to D^0 K^-) \over Br(B^-\to D^0\pi^-)}
     = 0.055 \pm 0.014 \pm 0.005
\eeq
which is consistent with the theoretical expectation.

The large background from $B^-\to D^0 \pi^-$ indicates a need for
a better particle identification. One way around it is to detect
the channel $B^-\to D^0 K^{*-}$, $K^{*-}\to K_S \pi^-$, where
the background from the Cabibbo-favored counterpart
$B^-\to D^0\rho^-$ is negligible. Such analysis is currently 
under way,

\section{$B\to$ charmless 2-body decays}

The interference of tree and penguin diagrams is expected to
cause substantial CP violating decay rate asymmetries in
charmless 2-body modes such as $B\to K\pi$ and $\pi\pi$.
Many methods have been proposed to extract the CKM phase
angles $\gamma$ and $\alpha$ using these modes.~\cite{Kpimethods}

In the rest frame of $\Upsilon 4S$, the $B$ mesons are nearly
at rest ($P_B\sim 350$ MeV/c), and thus
the light mesons are nearly monochromatic at $P\sim m_B/2$.
Such momentum is the highest that can be created by a $B$ decay,
and as a consequence,
the background from generic $B$ decays are small.
The main background thus comes from the continuum events
where the two-jet events $e^+e^-\to q	\bar q$ can generate
particles with momenta all the way up to the beam energy. 
And the continuum
background is suppressed by the same set of parameters as
in the $D^0 K^-$ analysis discussed above.

\begin{figure}[htb]
\vspace{-.1cm}
\centerline{\epsfig{figure=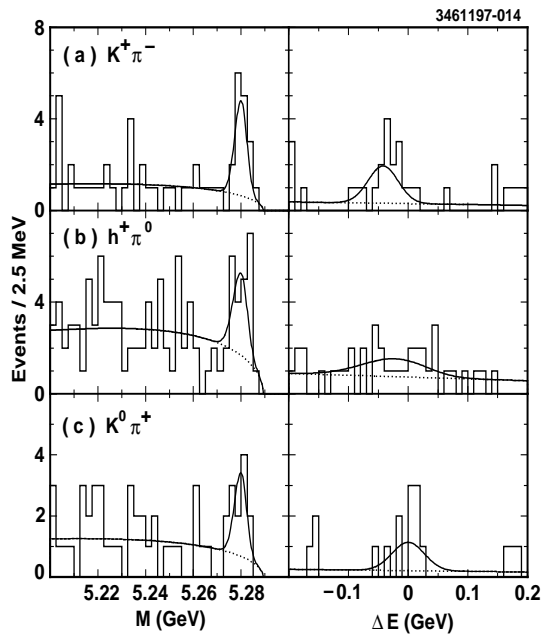,width=2.8in, height=3.3in}}
\caption{\label{fig:Kpi}
   The beam-constrained mass distribution for the modes $B^-\to K^-\pi^+$,
   $h^-\pi^0$, and $K_S \pi^-$ where
   $h^-$ is $K^-$ or $\pi^-$. All charged tracks are assumed to be
   pion. The dotted lines show the continuum backgrounds.}
\end{figure}

As in the case of $D^0 K^-$, it is critical to distinguish
charged kaons from charged pions and again we use the ionization loss
information ($dE/dx$) and the $B$ meson energy shift that occurs when
the light meson mass is miss-assigned.
Figure~\ref{fig:Kpi} shows the beam constrained mass
distribution for the modes $B^-\to K^-\pi^+$,
$h^-\pi^0$, $K_S \pi^-$,~\cite{CLEOKpiPRL} 
where $h^-$ is either $K^-$ or $\pi^-$.
The variable $\Delta E$ is the total energy of the candidate particles
minus the beam energy, and expected to peak at zero for the signal
when the masses are correctly assigned.
When the wrong mass assignment is made (namely, $K\lra\pi$ flip),
$\Delta E$ shifts by about 42 MeV. In the figure, the charged kaons
or charged pions are selected by the ionization loss only without 
using $\Delta E$. All charged tracks are assigned the pion mass
and thus one sees that the $\Delta E$ peak for $K^-\pi^+$ is
shifted. The $dE/dx$ also provides $\sim1.8\sigma$ separation between
$K^+$ and $\pi^+$ at $p=2.6$ GeV/c (2.0$\sigma$ for the new dataset
taken with a He-based drift chamber gas), and $\Delta E$ gives
close to $2\sigma$ separation for a single $K\lra \pi$ flip.
Final numbers are obtained by 
likelihood fit using the beam constrained mass, $\Delta E$,
$dE/dx$ and the set of variables used for the continuum
suppression.  Table~\ref{tb:Kpi}
gives the results where all numbers are published values~\cite{CLEOKpiPRL}
except for the column labeled `ICHEP98' which shows the
values updated at the ICHEP 1998 conference.~\cite{CLEOKpi98}
\begin{table}[hbt]
  \caption{\label{tb:Kpi} Measured branching fractions for $B\to \pi\pi$,
   $K\pi$, $KK$.}
  \begin{center} \footnotesize 
  \vskip 0.1in
  \begin{tabular}{l@{\tbst}rccc} 
    \hline\hline
      & $N$(signal)& signif. & $Br(10^{-5})$(PRL) &  $Br(10^{-5})$(ICHEP98)\\
     \hline
    $\pi^+\pi^-$ &  9.9  & $2.2\sigma$ & $<1.5$ & $<0.84$ \\
    $\pi^+\pi^0$ & 11.3  & $2.8\sigma$ & $<2.0$ & $<1.6$ \\
    $\pi^0\pi^0$ &  2.7  & $2.4\sigma$ & $<0.93$ & \\
     \hline
    $K^+\pi^-$   & 21.6  & $5.6\sigma$ & $1.5^{+0.5}_{-0.4} \pm 0.1 \pm 0.1$ 
                                & $1.4 \pm 0.3 \pm 0.2$ \\
    $K^+\pi^0$   &  8.7  & $2.7\sigma$ & $<1.6$ & $1.5 \pm 0.4 \pm 0.3$ \\
    $K^0\pi^+$   &  9.2  & $3.2\sigma$ & $2.3^{+1.1}_{-1.0} \pm 0.3 \pm 0.2$
                        & $1.4 \pm 0.5 \pm 0.2$ \\
    $K^0\pi^0$   &  4.1  & $2.2\sigma$ & $<4.1$ & \\
     \hline
    $K^+K^-$     &  0.0  & $0.0\sigma$ & $<0.43$ & $<0.24$ \\
    $K^+K^0$     &  0.6  & $0.2\sigma$ & $<2.1$  & $<0.93$ \\
    $K^0K^0$     &   0   &      -      & $<1.7$  & \\
    \hline
    $h^+ \pi^0$  &  20.0 & $5.5\sigma$ & $1.6^{+0.6}_{-0.5} \pm 0.3 \pm 0.2$ 
         & \\
    \hline\hline
  \end{tabular}
  \end{center}
\end{table}

One sees that we are starting to observe quite a few modes that are
important in measuring the CP violating CKM angles. Most of these modes
are self-tagging and do not require measurements of decay times
(including the neutral $B$ meson modes). Studies to measure decay
asymmetries of these and other modes are currently under way.

Table~\ref{tb:orare} lists the other 
hadronic rare modes of interest
that involve $\eta^{(\prime)}$, $\omega$, or vector mesons.
There are upper limits for other modes which can be found in
the references.~\cite{CLEOetaph,CLEOomegold}
\begin{table}[hbt]
  \caption{\label{tb:orare} Measured branching fractions for 
    modes with $\eta^{(\prime)}$ and $\omega$.}
  \begin{center} \footnotesize 
  \vskip 0.1in
  \begin{tabular}{l@{\tbst}ccc} 
    \hline\hline
      & $Br(10^{-5})$ & significance & ref. \\
     \hline
    $\eta'K^-$ &  $7.4^{+0.8}_{-1.3}\pm0.9$ & $12.7\sigma$
         & \cite{CLEOetapK} \\
    $\eta'K^0$ &  $5.9^{+1.8}_{-1.6}\pm0.9$ & $7.3\sigma$
         & \cite{CLEOetapK}  \\
    $\eta'\pi^-$  &  $<1.2$ &  & \cite{CLEOetapK} \\
    $\eta h^-$    &  $<0.8$ &  & \cite{CLEOetaph} \\
    $\eta'K^{*0}$ &  $<9.9$ &  & \cite{CLEOetaph} \\
    $\eta K^{*0}$ &  $<3.3$ &  & \cite{CLEOetaph} \\
    \hline
    $\omega K^-$     &  $1.5^{+0.7}_{-0.6}\pm0.2$ & $4.3\sigma$ 
         &  \cite{CLEOomeg} \\
    $\omega \pi^-$   &  $1.1^{+0.6}_{-0.5}\pm0.3$ & $2.9\sigma$ 
         &  \cite{CLEOomeg} \\
    $\phi K^*\;^\dag$   &  $1.3^{+0.7}_{-0.6}\pm0.2$     & $3.5\sigma$ 
         &  \cite{CLEOomeg} \\
    $\phi K^-$   &  $<0.53$     &     &  \cite{CLEOomeg} \\
    \hline
    $\rho^0 K^0$   &  $<3.6$ & & \cite{CLEOPV} \\
    $K^{*+}\pi^-$  &  $<4.5$ & & \cite{CLEOPV} \\
    $K^{*+}K^-$    &  $<1.1$ & & \cite{CLEOPV} \\
    \hline\hline
    \multicolumn{4}{l}{$^\dag$Average of $B^-\to\phi K^{*-}$ and
        $B^0\to\phi K^{*0}$.}
  \end{tabular}
  \end{center}
\end{table}
Some salient features one notices are:
(1) the $\eta'K^-$ rate is large, (2) $\eta'K^- > \eta K^-$, and
(3) $\eta' K \gsim \eta' K^*$.

Variety of mechanisms have been proposed to explain these features.
The large $\eta'K^-$ rate may be due to $c\bar c$ content of 
$\eta'$,~\cite{HaZhex} color-octet $c\bar c$ contribution,~\cite{ChTs}
new physics to enhance $b\to sg$,~\cite{HoTs,KaPe}
$b\to s gg$ followed by $gg\to \eta'$,~\cite{Ali+} and
gluon fusion where one gluon is from valence quarks.~\cite{Armady+,DuKimYang}
They all have something to do with gluon-$\eta'$ coupling, and many 
are closely related; in fact, when the gluon creation is dominated
by $c\bar c$ loop, it may be interpreted as $c\bar c$ content (or coupling)
to $\eta'$. Often, it is said that the 
$c\bar c$ content of $\eta'$ would result in $\eta'K^*/\eta'K \sim 2$
which is in contradiction with the experiment. This is not necessarily so.
With the value of the $c\bar c$ coupling needed to explain the $\eta'K$ rate, 
this ratio is still consistent with the data.~\cite{ChTs}
The observation $\eta'K^- > \eta K^-$ can be naturally explained
by the fact that gluon couples to $u\bar u$, $d\bar d$, and $s\bar s$
with the same amplitude, and the valence quark content of $\eta'$
and $\eta$ are roughly (ignoring the mixing)
$\eta' = u\bar u+d\bar d+s\bar s$ while $\eta = u\bar u + d\bar d - 2s\bar s$;
namely, $u\bar u+d\bar d$ and $s\bar s$ interfere constructively for
$\eta'$ and destructively for $\eta$.
The feature $\eta' K \gsim \eta' K^*$ is predicted by many of the
above mechanisms and agrees with the qualitative argument by
Lipkin.~\cite{Lipkin}
At the end of the day, however, one should note that the standard
4-quark operators (penguins included) can explain the observed
rates within the errors.~\cite{DaHePa}

\section{Inclusive $\eta'$ production}

As mentioned above, the gluonic penguin process ($b\to s g$) plays 
a critical role in direct CP violation in $B$ decays.
One signature of gluonic penguin is excess production
of $\eta'$ which is believed to be `gluon rich'.

\begin{figure}[htb]
\vspace{0.01cm}
\centerline{\epsfig{figure=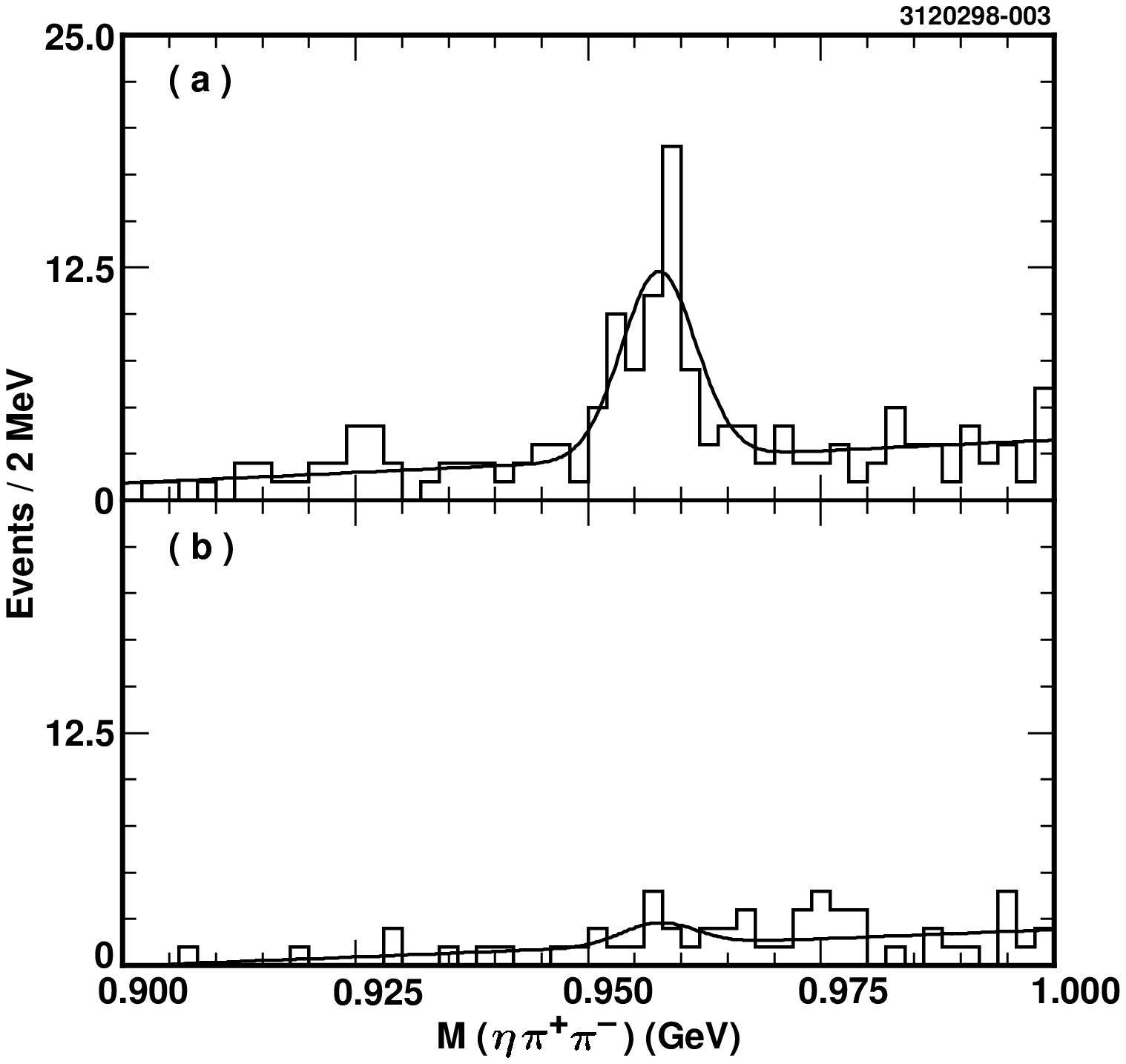,width=2.8in, height=2.8in}}
\vspace{-0.01cm}
\caption{\label{fig:Metap}
   The mass distribution for the $\eta'$ candidates for
   $2.0 < p_\eta' < 2.7$ GeV/c when there is at least one
   $B$ candidate in the signal region (see text).
   It is plotted for the on-resonance data (a) and for
   the continuum data (b).}
\end{figure}

In searching for $\eta'$ which might be originating from
$b\to s g$, one looks for them at the end point of 
the momentum spectrum in order to suppress $\eta'$ from $b\to c$ 
processes. The momentum window is chosen to be
$2.0 < p_{\eta'} < 2.7$ GeV/c.
The background from continuum is suppressed by
the pseudo-$B$-reconstruction technique which was used
previously~\cite{CLEObsgam} for the inclusive study of the $b\to s \gamma$
(electro-magnetic penguin) process.
Namely, we explicitly reconstruct $B\to \eta' K^\pm (n\pi)$ 
where $n\le4$ and at most one neutral pion can be included, 
then the invariant mass of the decay $\eta'\to \eta\pi^+\pi^-$ 
($\eta \to \gamma\gamma$) is plotted when there is at least 
one $B$ candidate in the signal region of $\Delta E$ vs. $M_B$.
The $\eta'$ mass distributions~\cite{CLEOetap} for the on-resonance 
data and for the continuum data are shown in Figure~\ref{fig:Metap}.
The effective luminosity for the continuum data relative to the
on-resonance data is 0.524; namely, the continuum background
in (a) is about twice the amount seen in (b). 
The excess in (a) after the continuum background is subtracted
is $39.0 \pm 11.6$.

The continuum-subtracted 
$\eta'$ signal is due to $B$ decays in general and not necessarily
due to $b\to s g$. Possible backgrounds are
\begin{enumerate}
  \item $b \to D,D_S \to \eta'$. The rates and spectra of $D$ and $D_S$
       in $B$ decay
       are well-measured and their decays to $\eta'$ are also experimentally
       known reliably. The contribution from these sources are estimated
       to be less than 0.2 events in the signal region.
  \item Color-allowed modes $b\to (DX) (\eta' X)$
       where the total charge of ($\eta'X$) is $-1$ (from $W^-$). 
       This contribution
       is modeled by 3-body decay $B\to D\eta'\pi$ with flat phase space 
       and searching for $\eta'$ at a lower momentum range. Upper limit
       is set at $<1.4$ events in the signal region.
  \item $b\to u \to \eta'$. Theoretical estimates~\cite{AtSo,ChTs}
       predicts that $3.5\sim7.0$\%\ of the signal is due to this
       source. The uncertainties, however, are large.
  \item Color-suppressed modes $B\to D^{(*)0}\eta'$. Theoretical
       branching fractions predictions for these 
       channels~\cite{Class2Th} leads to 
       $2.1\sim8.6$\%\ of the signal being due to this source albeit
       with large uncertainties. Experimentally, limits can be
       set by searching for $D^0$ or $D^{*0}$ in the signal sample
       which indicates that up to 41\%\ of the signal can be due to
       this source.
\end{enumerate}

About 10\%\ of the signal is the exclusive mode $\eta'K^-$ and 
for the rest the recoil mass distribution 
broadly peaks in the region 1.7 GeV to 2.4 GeV. 
The fall off at the high mass end is due to the cut 
$p_{\eta'}>2.0$ GeV/c. There is no signal seen for $\eta'K^*$. 
The branching ratio corresponding to the excess is 
\beq
   Br(B\to\eta' X_s) = (6.2\pm1.6\pm1.3^{+0.0}_{-1.5})\times10^{-4}
   \quad(2.0<p_{\eta'}<2.7 GeV)
\eeq
where the central number corresponds to the case where the backgrounds
discussed above are set to zero, and the lower value of the last
error corresponds to 1.5 times the worst case theoretical prediction
for the backgrounds discussed above.

If the excess is indeed largely due to $b\to s g$ process, the
measured $\eta'$ production is probably too large compared to
the conventional process where the corresponding 4-quark 
penguin operators are naively coupled
to $\eta'$ meson.~\cite{DaHePa}
Possible enhancement mechanisms proposed are related to those
proposed for the large exclusive $\eta'K$ rates: 
$c\bar c$ content of $\eta'$ which can
couple to the Cabibbo-favored transition $b\to c\bar c s$,~\cite{HaZh}
color-octet $c\bar c$ contribution,~\cite{YuCh}
coupling of $\eta'$ to two gluons through the QCD 
anomaly resulting in three-body decay $b\to \eta' s g$~\cite{AtSo}
or two-body decay $b\to \eta' s$,~\cite{Fritzsch} and new physics that
enhances $b\to s g$.~\cite{HoTs,KaPe} The 2-body decay tends to produce
hard $\eta'$'s (namely, small recoil mass) and seems at odd with the
measurement. At this point, there appears to be no strong 
evidence to force us to introduce new physics.

\section{Final-state-interaction phases}

Decay rate CP asymmetry in general requires interference of more than one
processes having different
final-state-interaction (FSI) phases as well as different weak phases.
In $B$ decays, final-state-interaction may be looked for in the
re-scattering $\bar B^0\to D^+\pi^- \to D^0\pi^0$ where $D^+\pi^-$
is color-favored while $D^0\pi^0$ is color-suppressed,
and an enhancement of $D^0\pi^0$ beyond what is expected from
naive factorization would indicate re-scattering.
Currently, however, only upper limit exists:~\cite{CLEOcl2}
$Br(\bar B^0\to D^0\pi^0) < 1.2\times10^{-4}$ while
factorization predicts around $1\times10^{-4}$. Thus,
no evidence for re-scattering is observed in this channel.

Another way is to search phase differences in helicity
amplitudes for $B\to VV$ decays such as $\Psi K^*$ and $D^*\rho$. 
Full angular fits has been performed for both 
modes. The analysis of 
$\Psi K^*$ modes~\cite{CLEOPsiKst} 
showed no evidence of FSI phase. The same analysis measured
the parity content of the $\Psi K^*$ final state to be
$16\pm 8\pm 4$\%\ parity $-$. Since the final state
$\Psi K^{*0}$ ($K^{*0}\to K_S\pi^0$) has C$=+$, the CP
is mostly $+$ which is opposite to the CP of the $\Psi K_S$
final state. With full angular analysis, this modes is equivalent
to about 1/4 of the $\Psi K_S$ sample in measuring the CKM angle
$\beta$.

The angular distribution for $B\to D^*\rho$ is given in terms of
the three helicity amplitudes $H_{\pm,0}$ by
\beqa
   {32\pi\over 9\Gamma}{d^3\Gamma\over d\cos\theta_1 d\cos\theta_2 d\chi}
   \hspace{-1.5in}&& \nonumber \\
   &=& 4|H_0|^2 \cos^2\theta_1\cos^2\theta_2 +
    (|H_+|^2 + |H_-|^2)\sin^2\theta_1\sin^2\theta_2 \nonumber \\
   && +[ \Re(H_-H^*_+) \cos2\chi + \Im(H_-H^*_+)\sin2\chi]
       2\sin^2\theta_1\sin^2\theta_2                \nonumber \\
   && +[\Re(H_-H^*_0 - H_+H^*_0)\cos\chi +
        \Im(H_-H^*_0 - H_+H^*_0)\sin\chi]\sin2\theta_1\sin2\theta_2
      \nonumber 
\eeqa
where $\theta_1$ ($\theta_2$) is the decay polar angle of
$D$ (charged $\pi$) in the $D^*$ ($\rho$) rest frame and $\chi$ is
the azimuthal angle between the two decay planes in the $B$ 
rest frame. Assuming $CP$ symmetry in the decay, helicity
amplitudes of $B$ and $\bar B$ are related by
$H_\lambda(\bar B) = H_{-\lambda}(B)$ ($\lambda = +,-,0$)~\cite{HelRel}
which corresponds to flipping the sign of $\chi$.
The data samples of $B$ and $\bar B$ are combined accordingly.
By convention, $H_0$ is taken to be real and the normalization is
$|H_+|^2 + |H_-|^2 + |H_0|^2 = 1$. Then,
the expression is invariant under the exchange $H_\pm\lra H^*_\mp$,
which means that the data cannot distinguish which is $H_+$ and 
which is $H_-$. The $V-A$ nature of the
interaction, however, indicates that $|H_+|<|H_-|$ for $\bar B$
and we thus define the larger of $|H_\pm|$ to be $H_-$
(opposite for $B$). Table~\ref{tb:dstrho} shows the
result of the fit.~\cite{CLEODstrho}
\begin{table}[hbt]
  \caption{\label{tb:dstrho} Helicity amplitudes for
     $\bar B^0\to D^{*+}\rho^-$ and $\bar B^-\to D^{*0}\rho^-$ .}
  \begin{center} \footnotesize 
  \vskip 0.1in
  \begin{tabular}{l@{\tbst}cccc} 
    \hline\hline
     & \multicolumn{2}{c}{$\bar B^0\to D^{*+}\rho^-$} &
       \multicolumn{2}{c}{$\bar B^-\to D^{*0}\rho^-$} \\
     & magnitude & phase & magnitude & phase \\
    \hline
      $H_0$ & $0.936$ & $0$ & $0.932$ & $0$ \\
      $H_-$ & $0.317\pm0.052\pm0.013$ & $0.19\pm0.23\pm0.14$ 
            & $0.283\pm0.068\pm0.039$ & $1.13\pm0.27\pm0.17$ \\
      $H_+$ & $0.152\pm0.058\pm0.037$ & $1.47\pm0.37\pm0.32$ 
            & $0.228\pm0.069\pm0.036$ & $0.95\pm0.31\pm0.19$ \\
    \hline\hline
  \end{tabular}
  \end{center}
\end{table}

There are signs of non-zero phases. Systematics checks have been
performed to confirm that it is not due to $\rho$ width, 
non-resonant $\pi\pi$ component, nor due to backgrounds.
The angular distribution expression above indicates that the
non-trivial phases results in $\sin\chi$ or $\sin2\chi$ distributions.
Thus, in principle such terms can be seen directly in 1-dimensional
distributions (with proper weighting as a function of $\theta_{1,2}$).
The data is unfortunately unable to demonstrate such signatures beyond
statistical errors. Alternatively, one could directly obtain the
coefficient of each term of the angular distribution by
the moment analysis.~\cite{moment} The result shows that
the coefficient $\Im(H_+H^*_-)$ is non-zero with $2.5\sigma$ significance.
The non-zero phases of Table~\ref{tb:dstrho} are more significant, and
it comes from the information contained in the 
relative values of the coefficients.

\section{Conclusions}

The ingredients required for the study of CP violation in B decays
are steadily accumulating and we are closing in on actually
measuring CP asymmetries. Many of the modes used in the measurements of
CKM angles $\alpha$ and $\gamma$ as well as
the `easy' $\beta$ are already observed,
and with a little more statistics we may even be able to see
direct CP violation. There are strong indications that the 
necessary ingredients for
such asymmetries - gluonic penguin and final state phases -
indeed exist.

\section*{Acknowledgments}

I thank Sandip Pakvasa and Tom Browder
for useful conversations relating to the topics presented.


\begin{thebibliography}{99}
\bibitem{lowPD}
  Since the angular correlation is
  too weak for $P_D<1.3$ GeV/c, there we use the separately-measured
  $Br(\bar B\to D,\bar D)$ and $Br(\bar B\to \ell^-X)$,
  where $D,\bar D$ and the leptons are in the given momentum
  windows, to separate $(D\ell)_{opp.B}$ and $(D\ell)_{same\,B}$
  assuming $(D\ell^+)_{same\,B} = 0$.
  See Ref 4 above.
\bibitem{HelRel}
  This is true to all orders under CP.
  Another relation found in the literature is 
  $H_\lambda(\bar B) = H^*_{-\lambda}(B)$ which is true to the first
  order in perturbation under CPT. Since we are concerned with FSI
  which is a higher order process, this CPT relation does not apply.
\end{thebibliography}
\end{document}